# Semi-Assisted Signal Authentication Based on Galileo ACAS

*Ignacio Fernandez-Hernandez[+], Simon Cancela[++], Rafael Terris-Gallego\*, José A. López-Salcedo\*, Gonzalo Seco-Granados\*, C. O'Driscoll[ζ], J. Winkel[ζ], A. Dalla Chiara\*\*, C. Sarto\*\*, Vincent Rijmen[±], Daniel Blonski[±±], Javier de Blas[¥]*

*European Commission[+], GMV Aerospace and Defence[++], Universidad Autónoma de Barcelona\*, RHEA[ζ], Qascom Srl\*\*, KU Leuven[±], ESA[±±], EUSPA[¥]*

**ABSTRACT**

A GNSS signal authentication concept named *semi-assisted* authentication is proposed. It is based on the re-encryption and publication of subsequences of some milliseconds duration from an already existing encrypted signal. Some seconds after the subsequences are transmitted in the signal-in-space, the signal broadcasts the key allowing to decrypt the sequences and the a-posteriori correlation at the receiver. The concept is particularized as Galileo Assisted Commercial Authentication Service, or ACAS, for Galileo E1B, with OSNMA used for the decryption keys, and E6C, assumed to be encrypted in the near future. This work proposes the ACAS cryptographic operations and a model for signal processing and authentication verification. Semi-assisted authentication can be provided without any modification to the signal plan of an existing GNSS, without the disclosure of signal encryption keys, and for several days of autonomy, depending on the receiver storage capabilities.

**INTRODUCTION**

Global Navigation Satellite Systems (GNSS) location is based on two main inputs: on the one hand the satellite positions and time data, and on the other hand the pseudorange measurements obtained at the user receiver. GNSS data authentication is already in the signal in space for Galileo through OSNMA (Open Service Navigation Message Authentication) [1] and possibly other GNSS in the future. Pseudoranges are more difficult to authenticate, as they are generated in the receiver. However, as they are based on the GNSS signal spreading codes, authenticating the spreading code sequence provides an additional level of protection. The reason is that DS-SS (Direct Sequence – Spread Spectrum) techniques bury the signals below the noise floor in the receiver, so it difficult for an adversary to estimate the chips if not provided by the system independently. This is why spreading code total or partial encryption is proposed as the most effective way to authenticate GNSS signals and measurements. Other methods can use signal unpredictability from NMA cryptographic bits or symbols [2], vestigial signal monitoring [3], or consistency checks [4].

Civil SCA (Spreading Code Authentication) was proposed almost two decades ago [5] and can be implemented in many ways, many of which are recollected in [6]. In standalone mode, i.e. without any assistance, it is based on watermarks in the spreading code that are correlated a posteriori, when a decrypting key is disclosed [7]. If the receiver is assisted, i.e. it has a network connection, it can capture an encrypted signal snapshot and have it authenticated in a server. This concept has been already particularized for Galileo E1-E6 in [8]. The server may, or may not [9], know the decryption key. Another approach is that the receiver receives from the server a posteriori the correlation replica [10]. In these approaches, the receiver connects to the server for each authenticated position. Both standalone and assisted methods are under test for GPS Navigation Technology Satellite (NTS)-3 as CHIMERA [11]. Dedicated schemes for Galileo range authentication have also been proposed in the literature [12] and are under definition for Galileo's 2[nd] Generation.

The proposed scheme, here named *semi-assisted authentication*, is intended for Galileo users but it can be extrapolated to other systems. The GNSS has to transmit NMA, or at least unpredictable bits regularly, and an encrypted signal, e.g. a military or commercial one. The purpose is to maximize receiver autonomy to several days, yet allowing regular spreading code authentications.

Once the Galileo program officially confirmed in 2017 its intention to provide signal authentication [13], this concept [14] was proposed to alleviate receiver requirements yet providing SCA based on the already existing signal plan. This concept is applied to Galileo through the E1B OSNMA and the E6C pilot signal, soon to be encrypted.

After this introduction, we present our motivation and the general concept. Then, we describe the proposed scheme in detail, and how it is particularized for the Galileo system, including cryptographic operations, definition of the E6C snapshot, some signal processing parameters, and a proposed signal authentication verification check. We finalize with the conclusions and further work.

**MOTIVATION AND GENERAL CONCEPT**

The main motivation is to provide SCA by a GNSS system that transmits a spreading-code-encrypted signal and NMA, as Galileo, without changing the signal plan, satellite payload or key management scheme. Currently, all GNSS transmit encrypted signals: GPS P(Y) and M-code, Galileo PRS and CS (E6C), and so on. Galileo already provides NMA, but other GNSS may too do so in the future. If the SCA scheme can be designed from scratch without these constraints, other schemes may be more optimal, such as [11] or [12].

The service operator must know a priori the key used to encrypt the encrypted signal spreading code (or at least the keystream for those subsequences that are to be re-encrypted), and the future NMA keys transmitted in an open signal. These NMA keys are unpredictable to the user. For example, a TESLA chain like OSNMA's [15] fits this purpose. With these two elements, the operator can re-encrypt parts of the encrypted signal, at known instants, with the NMA cryptographic material to be later disclosed, to create *re-encrypted code sequences* (RECS), and publish them in advance.

The receiver downloads the RECSs, with their start and end times, for the desired autonomy period. At, or around, those times, the receiver records an Intermediate Frequency (IF) or baseband signal snapshot that should contain the encrypted signal, then waits for the NMA key to be disclosed, decrypts the RECS to obtain the original Encrypted Code Sequence (ECS), and correlates a-posteriori the signal snapshot with the ECS. If correlation occurs, the presence of the ECS is detected, and under certain conditions and hypotheses the receiver can assume that the probability that the signal is authentic is high. A key assumption is that the receiver has a loose synchronization time source, sufficient to prevent an adversary to replay the ECS after the NMA key is disclosed. This requirement is already in place for Galileo OSNMA.

The concept is depicted in Figure 1 for the system side and Figure 2 for the receiver side. Figure 1 shows an encrypting key $K_{encr}$ generating a keystream. The keystream will be multiplied with the GNSS spreading codes, encrypting the signal. Some sequences of the encrypted signal $ECS_j$, $ECS_{j+1}$… are selected for re-encryption at predefined slots. These sequences are then re-encrypted with an NMA key $K_j$, $K_{j+1}$,…, already known to the system operator, that will be publicly disclosed in the NMA bitstream with a certain delay after the ECS time. The sequences are encrypted with the NMA keys, forming $RECS_j$, $RECS_{j+1}$,…, and then these RECS, together with their associated time tags and other metadata are published in a server, for a certain period that can be of several days or even weeks, depending on the system operation and key management. In Figure 2, the receiver stores a signal snapshot containing the ECS, waits until the related NMA key ($K_j$) is disclosed, decrypts $RECS_j$, and performs the correlation, obtaining the de-spreading gain if the signal is found. We must highlight that this is a proposal that provides a confidence level on the pseudorange authentication but, in order to increase this confidence level, further receiver measures must be implemented. These may include power level checks such as AGC (Automatic Gain Control) monitoring [16], dedicated signal correlation tests such as [17], or search for vestigial signals [3]. While relevant, these measures are left out of the scope of this work, which just focuses on the RECS measurement generation.

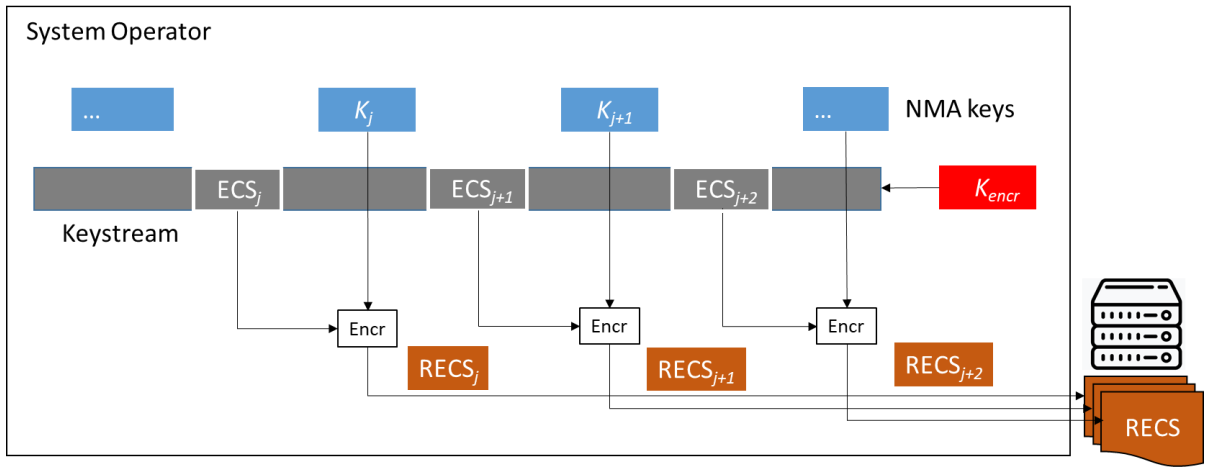

Figure 1 – Semi-assisted spreading code authentication – system side

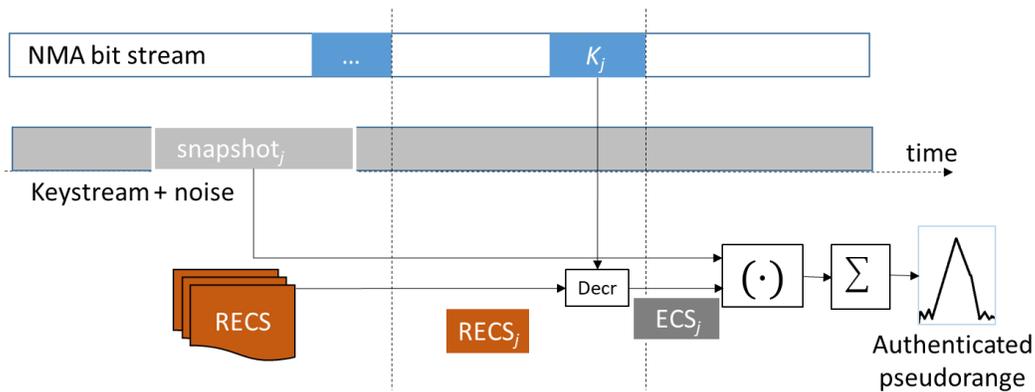

Figure 2 – Semi-assisted spreading code authentication – receiver side

## PARTICULARIZATION FOR GALILEO ACAS

This section describes the concept in detail and particularizes it to Galileo E1B/E6C signals. The concept is currently under implementation and testing as Galileo Assisted Commercial Authentication Service, or ACAS. An early version was also prototyped in the EC NACSET project [18]. The parameters here proposed may vary in the final operational specification of ACAS.

ACAS is based on two already existing Galileo signals: the Galileo E1B [19] including OSNMA [20], and the Galileo E6C pilot signal [21] providing the encrypted signal. This signal is currently unencrypted, but it will soon be encrypted to provide Galileo ACAS. The semi-assisted authentication in Galileo through ACAS is depicted in Figure 3.

A high-level concept of operations is proposed in the following steps:

1. The publication, in a server at the European GNSS Service Centre (GSC), of E6C RECS files encrypted with OSNMA-based keys. The server also provides authenticated satellite broadcast group delays (BGDs) between E1 and E6, as this information is not in the broadcast I/NAV message and therefore not authenticated by OSNMA. Note that Galileo HAS (High Accuracy Service) provides these BGDs but they are unauthenticated at the moment.
2. The RECS/BGDs, or a subset of them, are downloaded from the server and stored in the receiver.
3. The receiver starts up, acquires the Galileo E1B signal, and synchronizes with it and obtaining a data-authenticated PVT.
4. At the time RECS are expected, the receiver records a snapshot of E6C samples and stores it for later use.
5. After some time, the E1 I/NAV OSNMA key is received.
6. For each satellite, the RECS is decrypted with an OSNMA-based key, obtaining the ECS.
7. The ECS and the E6C snapshot are correlated. If the correlation is successful, an E6C pseudorange is generated.
8. E6C pseudoranges together with I/NAV authenticated data and E6 BGDs are used for a spreading code-authenticated position calculation.

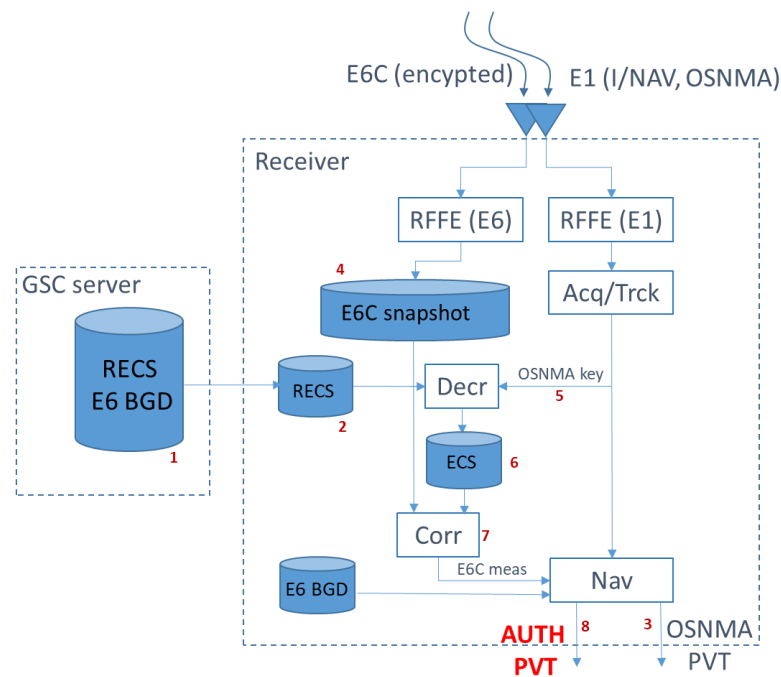

*Figure 3 – Semi-assisted CAS scheme, including the steps to deliver the service.*

**RECS and BGD files**

In addition to the sequences themselves, a RECS file downloaded from the server starts with a header that defines the start time, duration, satellite ID, RECS period $\tau_{RECS}$, or time between RECS, and length, $N_{chips}$, or number of chips per RECS. The RECS file header provides three additional parameters: a RECS Offset, which defines the position of the *start* of the RECS with respect to the start of an integer GST second, a

SLRECS (Slow RECS) Offset, which defines whether the RECS belongs to the current OSNMA key to be provided within the current (SLRECS=0), or the next 30-second OSNMA block (SLRECS = 1), or a later (SLRECS = 2, 3, ...) 30-second OSNMA block, and $\Delta\tau_{MAX}$, or RECS maximum random delay, which is a parameter to optionally randomize the position of each RECS. The file content authenticity can be assured by a digital signature. RECS parameters are depicted in Figure 4. The top plot shows no time randomization ($\Delta\tau_{MAX} = 0$), and the bottom plot shows randomization in the RECS start ($\Delta\tau \leq \Delta\tau_{MAX} \neq 0$). The receiver can determine $\Delta\tau$ only after the OSNMA key of the related period is disclosed.

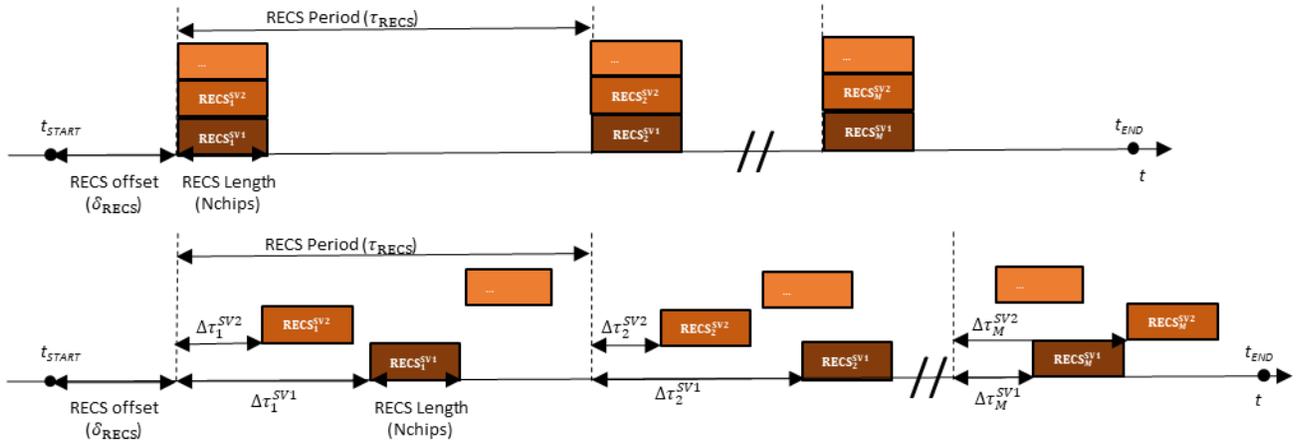

Figure 4 – Main parameters of RECS files, without randomization (top, $\Delta\tau_{MAX} = 0$), and with randomization (bottom, $\Delta\tau_{MAX} \neq 0$), for a total amount of M RECS

Regarding the BGD files, the system can predict the E1 I/NAV-E6C BGDs and provide them in separate files. Additional information may include the accuracy estimation over time, so its related uncertainty can be accounted for. The BGDs are very stable and in nominal conditions can be stably predicted for several days.

**Cryptographic Operations**

The cryptographic operations required at the receiver are the generation of the RECS decryption key, the generation of the randomization parameter, and the RECS decryption. Other possible operations, such as the digital signature verification of the RECS/BGD files can be based on existing standards and are not described here. The RECS decryption process for several OSNMA blocks is depicted in Figure 5, where two ECS/RECS per 30-second OSNMA block are depicted, and SLRECS=1.

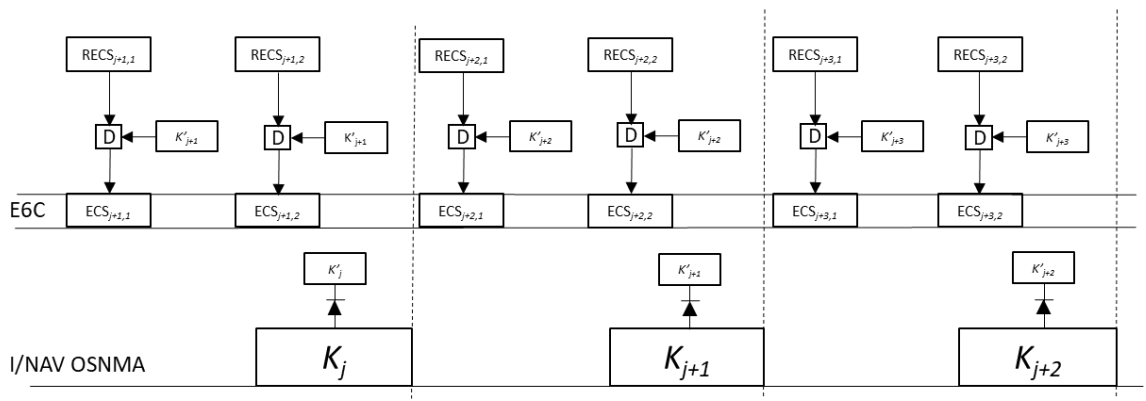

Figure 5 – RECS decryption process (SLRECS Offset = 1 and two RECS per 30-second OSNMA period)

RECS decryption key: Once the OSNMA key $K_j$ belonging to MACK (MAC and key) block $j$ is received and verified by the OSNMA keychain, the RECS decryption key $K'_j$ is generated as follows:

$$K'_j = \text{SHA256}(K_j) \tag{1}$$

where SHA256() is the hash function SHA-256 as per [22]. As SHA256 is a one-way function, this allows RECS decryption only when the OSNMA related key is disclosed, but, at the same time, the OSNMA keychain itself is not used at the moment of RECS encryption, minimizing OSNMA and ACAS key management dependencies. Therefore, ACAS and OSNMA operation are independent provided that, when the OSNMA keychain is generated, a parallel set $K'_j$ chain is generated as well.

RECS Time Randomization: The purpose of RECS time randomization is to avoid that all RECS are placed deterministically at a certain time in the signal. It is bounded by the parameter $\Delta\tau_{MAX}$, the maximum RECS random delay. If $\Delta\tau_{MAX} = 0$ there is no randomization, and the RECSs are deterministically located for each RECS period, as per Figure 4, top. If $\Delta\tau_{MAX} \neq 0$, RECS locations are delayed by a random time offset $\Delta\tau$ between zero and $\Delta\tau_{MAX}$. For example, if $\Delta\tau_{MAX} = 3$, RECS can be located with a delay which is a multiple of 8ms (or two E1B/C codes), i.e. 0, 8, 16 or 24 ms for $\Delta\tau = 0, 1, 2$ or 3, respectively, as per (6). The time offsets are generated from the OSNMA derived key ($K'_{j+\text{SLRECS}}$). As there is only one OSNMA key for every 30-second period, but more offsets may be required for each satellite per RECS period, we first use the AES cypher, as shown in Figure 6, to generate a sufficiently large cyphertext from which the random time offsets are extracted. In particular, the cyphertext is generated as follows:

$$(C_1,..,C_N) = \text{AES256}_{OFB}(K'_{j+\text{SLRECS}}, 0, \text{IV}) \tag{2}$$

where $(C_1,..,C_N)$ is the cyphertext block array from which the $\Delta\tau$s are generated, consisting of $N$ 128-bit blocks; $\text{AES256}_{OFB}(a, b, \text{IV})$ is the AES cipher in OFB (Output Feedback) mode and configured for 256-bit keys (AES-256), where $a$ is the key, $b$ is the plaintext, and IV is the initialisation vector, as per [23]. The IV used for $\text{AES256}_{OFB}$ is generated as follows:

$$\text{IV} = trunc(128, \text{SHA256}(P_j)) \tag{3}$$

Where $trunc(n, p)$ is the truncation function that retains the $n$ MSBs of the input $p$, and $P_j$ is a 128-bit plaintext generated as follows:

$$P_j = (GST_{SF,j}|p1) \tag{4}$$

where $GST_{SF,j}$ is the 32-bit GST associated to the OSNMA key $K_j$ as per [19]; and $p1$ is a padding array with 96 zeros. Then, the cyphertext block array is allocated to the random time offsets $\Delta\tau$s as follows:

$$C_1 = [B_1^1, B_1^2, \ldots, B_1^{16}], C_2 = [B_1^{17}, B_1^{18}, \ldots, B_1^{32}], C_3 = [B_1^{33}, B_1^{34}, \ldots, B_1^{48}], \tag{5}$$

$$C_4 = [B_2^1, B_2^2, \ldots, B_2^{16}], C_5 = [B_2^{17}, B_2^{18}, \ldots, B_2^{32}], C_6 = [B_2^{33}, B_2^{34}, \ldots, B_2^{48}],$$
$$C_7 = [B_3^1, B_3^2, \ldots, B_3^{16}], C_8 = [B_3^{17}, B_3^{18}, \ldots, B_3^{32}], C_9 = [B_3^{33}, B_3^{34}, \ldots, B_3^{48}],$$
$$C_{10} = [B_4^1, B_4^2, \ldots, B_4^{16}], C_8 = [B_4^{17}, B_4^{18}, \ldots, B_4^{32}], \ldots$$

where $B_i^k$ is a byte defining $\Delta\tau_i^k$, expressed as an integer number of 8-millisecond periods, as follows:

$$\Delta\tau_i^k = B_i^k \bmod (\Delta\tau_{MAX} + 1) \tag{6}$$

where mod is the *modulo* operator; $\Delta\tau_{MAX}$ is described in the RECS file; $i$ is the RECS position in 30-second period $j$ ($j$ omitted from (6) for simplicity); and $k$ is the SVID index, defining the satellite, and where only values from 1 to 36 are currently used. For example, if $\Delta\tau_{MAX} = 3$ and $B_2^1 = 5$, then $\Delta\tau_2^1 = 1$ (8 ms), which relates to the second RECS of the period for SVID=1.

Note that, while the randomization can protect against some denial of service attacks by making the part of the signal used for correlation less predictable, it can also make the signal snapshot bigger, therefore increasing the receiver storage requirements.

Figure 6 shows how to generate $C$ based on AES in OFB mode. The figure is based on the AES modes of operation as described in [23].

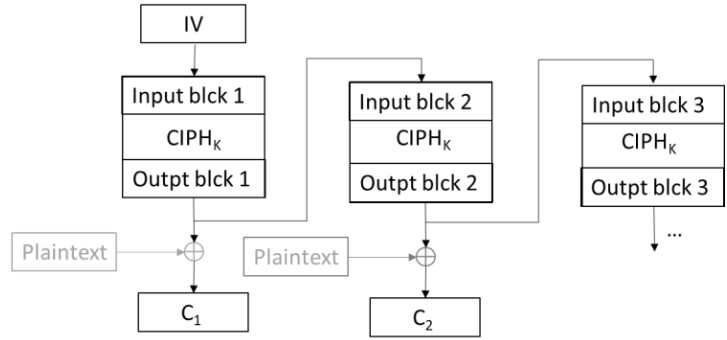

*Figure 6 – Generation of C for RECS randomization with AES cipher (CIPH) in OFB mode. The plaintext is depicted in grey as it is set to zero.*

RECS decryption: Finally, the ECS decryption process, particularised for the decryption key $K'_{j+\text{SLRECS}}$, taking into account the delay introduced by the *Slow RECS* offset, SLRECS, can also be performed by using AES. The Initialisation Vector (IV) for the cipher, applicable to all decryptions in a given RECS file, is determined as per (3). Then, for every satellite $k$:

$$\text{ECS}_{j,i} = \text{AES256}_{CBC}^{-1}(K'_{j+\text{SLRECS}}, \text{RECS}_{j,i}, \text{IV}) \tag{7}$$

where $\text{ECS}_{j,i}$ is the decrypted sequence and $\text{AES256}_{CBC}^{-1}(a, b, \text{IV})$ is the inverse cipher AES, configured for 256-bit keys (AES-256) in CBC mode to decrypt a plaintext $b$ with 256-bit key $a$. Note that the $j$ index defines the decrypting key, and the $i$ index allows for more than one RECS per decrypting key. Note also that RECS$_{j,i}$ must be an integer number of blocks of 128 bits, as required by the proposed AES implementation [24], configured to operate in CBC mode. Figure 7 shows how to generate the ECS based on AES in CBC mode. The figure is based on the AES modes of operation as described in [23].

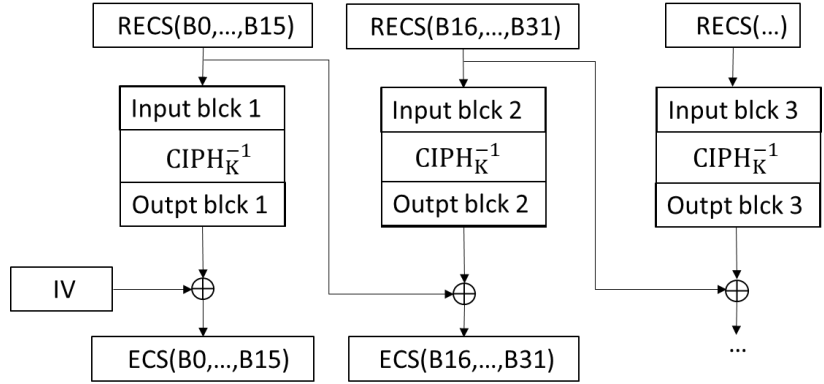

*Figure 7 – Generation of ECS from RECS and AES256 in CBC mode. RECS/ECS are represented in 16-byte arrays (e.g. B0,…,B15).*

**ECS and Snapshot Signal Correlation**

This section presents an implementation of the ACAS signal correlation process. This concrete implementation is dependent on prior E1 acquisition and is one of the possible approaches. The ACAS signal correlation process is generalized and further discussed in [25], including some simulated results. This reference also proposes other acquisition methods independent from E1.

E6C snapshot definition: We assume the E1 signal with OSNMA is trustable a priori and verify this hypothesis with the E6C ECS correlation. Therefore, at the time of the capturing the snapshot, the receiver is synchronized with GST through E1. Then, for a certain satellite $k$ and $ECS_j^k$, associated to a GST second $GST_j$, the a priori E6 snapshot start time $t_{start,j}^k$ referred to an absolute time reference GST, is defined as

$$t_{start,j}^k = GST_j + \delta_{RECS} + \tau_{prop}^k - \delta t_{sat,E1}^k + \delta t_{rx,E1} + \delta_{E1,E6}^k \quad (8)$$

Where $\delta_{RECS}$ is the RECS offset, $\tau_{prop}^k$ is the estimated propagation time of the satellite (typically between 77 ms and 97 ms for Galileo), $\delta t_{sat}^k$ is the satellite clock offset, at the time of the snapshot, in the E1 signal, which is known at the receiver, $\delta t_{rx,E1}$ is the receiver clock bias, also estimated by the receiver, based on an E1 signal solution, and $\delta_{E1,E6}^k$ is the estimated time bias between the E1 and E6 signals, detailed later. $\delta_{E1,E6}^k$ ensures that the ECS is fully contained in the snapshot. Note that we drop the index $i$ for simplicity but without loss of generality; the proposal can be understood as for the case of $i=1$, i.e. one ECS per OSNMA key, but it can be generalized to $i>1$. Note also that our definition of $t_{start,j}^k$ neglects the errors in the estimations of its related parameters, which may be in the order of a few meters and therefore usually below one E6C chip (~30m).

Then, the snapshot end time $t_{end,j}^k$ can be defined as

$$t^k_{end,j} = t^k_{start,j} + \Delta\tau_{MAX} + \frac{Nchips}{R_c} \tag{9}$$

where $\Delta\tau_{MAX}$ is the maximum random offset, $Nchips$ is the RECS length, in chips, and $R_c$ is the chip rate ($R_c = 5.115 \cdot 10^6$ chips per second for E6C). A snapshot that covers the time interval $t^k_{start,j} \leq t < t^k_{end,j}$ usually covers all chips in the RECS for satellite $k$ in epoch $j$. Again, these times are referred to GST. If the receiver captures only one snapshot for all satellites $k=1..\kappa$, then the snapshot needs to be broadened by some tens of ms, as defined by $t^{all}_{start,j}$ and $t^{all}_{end,j}$:

$$t^{all}_{start,j} = GST_j + \delta_{RECS} + \min(\boldsymbol{\tau}^{all}_{sat,prop}) + \delta t_{rx,E1} \tag{10}$$

$$t^{all}_{end,j} = GST_j + \delta_{RECS} + \max(\boldsymbol{\tau}^{all}_{sat,prop}) + \delta t_{rx,E1} + \Delta\tau_{MAX} + \frac{Nchips}{R_c} \tag{11}$$

Where max() and min() are the maximum and minimum operators and $\boldsymbol{\tau}^{all}_{sat,prop}$ is a vector with propagation times and signal time offsets:

$$\boldsymbol{\tau}^{all}_{sat,prop} = (\tau^1_{prop} - \delta t^1_{sat,E1} + \delta^1_{E1,E6}, \ldots, \tau^\kappa_{prop} - \delta t^\kappa_{sat,E1} + \delta^\kappa_{E1,E6}) \tag{12}$$

This guarantees that the snapshot starts at the earliest-to-arrive satellite, and finishes at the end of the ECS of the latest satellite, considering both propagation, satellite clock, and biases.

When $K'_{j+SLRECS}$ is determined, the random times $\Delta\tau^k_j$ (index $i$ is dropped without loss of generality) are calculated as per (6). Then, the snapshot of samples for each satellite $j$ can be narrowed as follows:

$$t^k_{acq-start,j} = t^k_{start,j} + \Delta\tau^k_j \tag{13}$$

$$t^k_{acq-end,j} = t^k_{acq-start,j} + \frac{Nchips}{R_c} \tag{14}$$

<u>E1-E6 time and frequency offsets</u>: The term $\delta^k_{E1,E6}$ models the time bias estimation between the E1 and E6 pseudoranges. It allows estimating the E6 pseudorange from the already measured (or estimated) E1 measurement. It includes all the time offsets that are different between signals. It is modelled as

$$\delta^k_{E1,E6} = \widehat{BGD}^k_{sat,E1,E6} + \delta I_{E1,E6} + \widehat{HWB}_{rx,E1,E6} \tag{15}$$

with the following definitions:

- $\widehat{BGD}^k_{sat,E1,E6}$ is the *estimation* (denoted as $\widehat{(\cdot)}$) of the satellite bias, or broadcast group delay (BGD) between E1B/C and E6C signals transmitted by satellite *k*. In particular:

$$\delta t^k_{sat,E6} = \delta t^k_{sat,E1} - \widehat{BGD}^k_{sat,E1,E6} \qquad (16)$$

where $\delta t^k_{sat,E6}$ is the satellite time offset as observed in the E6 signal. $\widehat{BGD}^k_{sat,E1,E6}$ is an input for ACAS from the ACAS BGD file, as shown in Figure 3. Galileo E1-E6 BGDs are usually in the order of up to a few meters and, while the BGD files are generated a priori, the BGDs are expected to be stable, allowing the correlation process.
- $\widehat{HWB}_{rx,E1,E6}$ is the receiver hardware bias between E1B/C and E6C, to be added to the receiver bias in E1, $\delta t_{rx,E1}$. If the E6 sample-grabbing hardware is different than the E1B/C tracking one [26], hardware biases may be higher than those in standard multi-frequency receivers. $\widehat{HWB}_{rx,E1,E6}$ should be calibrated in the receiver and estimated a priori. An example of real biases in a prototype device compatible with ACAS is available in [26].
- Finally, the term $\delta I_{E1,E6}$ is an additional offset due to the frequency dispersive effect of the ionosphere, which deserves particular attention and is treated in the following subsection.

E1 vs. E6 ionospheric delay: We define the pseudorange equation for E1, in the time domain, and after applying the atmospheric correction models, as follows:

$$\tau^k_{E1} = \frac{r^k}{c} + \delta t_{rx,E1} - \delta t^k_{sat,E1} + \epsilon_{I,E1} + \epsilon_T + \epsilon_{MP,E1} + \epsilon_{n,E1} \qquad (17)$$

where $r^k$ is the receiver-satellite *k* range, *c* is the light speed, $\epsilon_{I,E1}$ is the remaining ionospheric error for the E1 measurement, after applying an ionospheric model, $\epsilon_T$ is the tropospheric error after applying a tropospheric model, $\epsilon_{MP,E1}$ is the multipath error in E1 and $\epsilon_{n,E1}$ is the receiver noise error also for the E1 measurement. The total ionospheric error in E1, before correction, can be defined as follows:

$$I_{E1} \approx \frac{40.3 \text{TEC}}{f_1^2} = \hat{I}_{E1} + \epsilon_{I,E1} \qquad (18)$$

where TEC is the total electron content, $\hat{I}_{E1}$ is the corrected part from the ionospheric model, and $\epsilon_{I,E1}$ is the non-corrected part appearing in (17).

We assume that the ionospheric error is estimated from a given ionospheric model, such as Galileo's NequickG [27] or GPS's Klobuchar [28], which corrects a certain percentage of the ionospheric error. e.g. the Klobuchar model is claimed to correct about 50% of the ionospheric error in average ionosphere conditions. NequickG is preferred as its coefficients are authenticated by Galileo OSNMA.

Then, the term $\delta I_{E1,E6}$ needs to account for the fact that, due to the frequency dispersion in the ionosphere, the error in E6 is higher, $I_{E6} \approx \frac{40.3 \text{TEC}}{f_6^2}$, so $\hat{I}_{E6}$ must be increased accordingly. Therefore, in the absence of a more refined model, a user can estimate $\delta I_{E1,E6}$ as follows, where $f_1 = 1575.42$ MHz and E1 $f_6 = 1278.75$ MHz for E1 and E6 respectively:

$$\hat{I}_{E6} \approx \frac{40.3 \text{TEC}}{f_6^2} = \frac{40.3 \text{TEC}}{f_1^2} \cdot \frac{f_1^2}{f_6^2} = \hat{I}_{E1} \cdot \frac{f_1^2}{f_6^2} = \hat{I}_{E1} + \delta I_{E1,E6} \quad (19)$$

$$\delta I_{E1,E6} = \hat{I}_{E1}\left(\frac{f_1^2}{f_6^2} - 1\right)$$

E1 vs E6 Frequency Offset: Regarding the frequency domain, the carrier frequency of the signal as processed by the receiver is mainly affected by the relative satellite-receiver velocity (Doppler effect) and the receiver clock oscillator drift. This adds a different frequency offset in the E1 and E6 bands, as the frequency carriers have different wavelengths, and can be modelled as:

$$\widehat{\Delta f}_{j,E6}^k = \Delta f_{j,E1}^k \cdot \frac{\lambda_1}{\lambda_6} = \Delta f_{j,E1}^k \cdot \frac{f_6}{f_1} \quad (20)$$

where $\widehat{\Delta f}_{j,E6}^k$ and $\Delta f_{j,E1}^k$ are the frequency deviations in the E6 and E1 bands, and $\lambda_1$ and $\lambda_6$ are the carrier wavelengths ($\lambda_1 = c/f_1$; $\lambda_6 = c/f_6$). The frequency offset can be expressed for convenience as $\widehat{\Delta f}_{j,E6}^k = \Delta f_{j,E1}^k + \Delta f_{E1,E6}^k$, where $\Delta f_{E1,E6}^k = \Delta f_{E1}^k(1 - \frac{\lambda_1}{\lambda_6})$.

E6C Signal Correlation and Measurement Generation: The receiver can use as an input the code phase and frequency estimations, $\tau_{j,E1}^k$ and $f_{j,E1}^k$ respectively, from E1, for a given satellite $k$ and instant identified by $j$ and, where $\tau_{j,E1}^k$ can be defined as the start of the 4-ms E1 spreading code, which is synchronized with the start of the ECS sequence in E6C, $\hat{\tau}_{j,E6}^k$, except for $\delta_{E1,E6}^k$:

$$\hat{\tau}_{j,E6}^k = \tau_{j,E1}^k + \delta_{E1,E6}^k \quad (21)$$

With $\hat{\tau}_{j,E6}^k$ and $\widehat{\Delta f}_{j,E6}^k$ the receiver can acquire the ECS in snapshot mode by correlating the ECS replica modulated with $\widehat{\Delta f}_{j,E6}^k$ with the snapshot. If the E6C encrypted signal is present in the snapshot, at the end of this step the receiver obtains a pseudorange and Doppler frequency measurement in E6C, synchronized with E1 (except for noise and biases), which can be verified for the position authentication. The process is depicted in Figure 8. In this figure, the subindex $j$ is also omitted for simplicity. The actual pseudorange (in the time domain) and frequency Doppler measurements from the ECS correlation are represented as $\tau_{E6}^k$ and $\Delta f_{E6}^k$. They are expected to be in the uncertainty window defined in the Figure by its maximum values $F_{max}^k$ and $T_{max}^k$. This uncertainty window is expected to be very small in the frequency domain, so if the E1 signal frequency is well tracked, the frequency error may be neglected. In the time domain, $T_{max}^k$ may require correlation across one to a few chips, mainly depending on ionosphere and multipath errors. Another practical aspect for consideration is that the receiver may not generate a pseudorange measurement exactly synchronized with $\tau_{j,E1}^k$, for example, if measurements are generated at a 50-Hz rate and there is one measurement every 20 ms. In this case, the receiver can make an estimation of $\tau_{j,E1}^k$ based on the closest measurement the known range rate. These aspects will be characterized in detail as part of further work for different receiver cases.

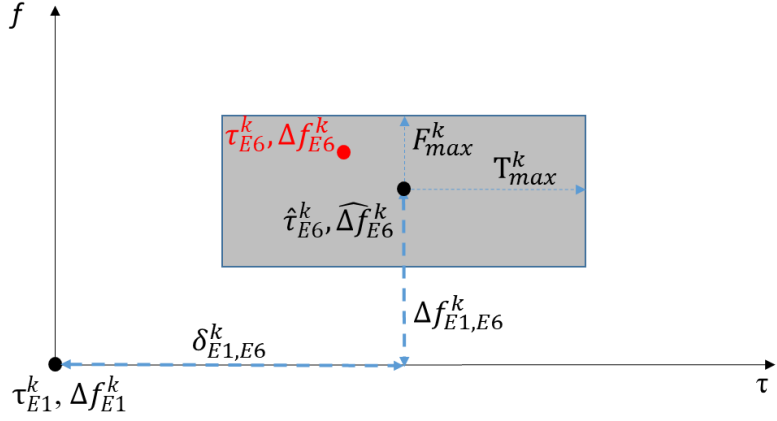

Figure 8 – E6C pseudorange and Doppler frequency ($\tau_{E6}^k$, $\Delta f_{E6}^k$) in from E1 measurements ($\tau_{E1}^k$, $\Delta f_{E1}^k$).

<u>Position Authentication Check based on E6 Measurements</u>: There are several ways by which the authentication verification can be performed. Aspects like the E6C vs E1 processing gain, or the difference in E6C and E1Based positions, can be assessed. For simplicity, here we propose a model based on the difference between $\tau_{j,E6}^k$ and $\hat{\tau}_{j,E6}^k$. The difference between the measured and estimated E6C pseudorange is compared with a threshold, defined by an upper bound of the estimated measurement error, as already outlined in [29]. For a given verification $ECS_j^k$ associated to measurement $\tau_{j,E6}^k$, we define a boolean variable $\xi_j^k$ and perform the following verification:

$$\left|\tau_{j,E6}^k - \hat{\tau}_{j,E6}^k\right| \leq \gamma_{auth} \to \xi_j^k = 1 \; (measurement\; authenticated) \quad (22)$$

$$\left|\tau_{j,E6}^k - \hat{\tau}_{j,E6}^k\right| > \gamma_{auth} \to \xi_j^k = 0 \; (measurement\; not\; authenticated)$$

$$\prod_{k=1}^{\kappa} \xi_j^k = 1 \to E1\; I/NAV_j\; position\; authenticated \quad (23)$$

We define the authentication threshold with the assumption that, in the absence of attacks, i.e., in nominal conditions, the error contributions follow zero-mean normal distributions, and therefore the overall error of the test variable also follows a zero-mean normal distribution characterized by a standard deviation $\sigma_{auth}$. Then,

$$\gamma_{auth} = K\sigma_{auth} \quad (24)$$

The value of $K$ can be defined for the desired level of confidence (e.g. $K = 2$ for a 95% or $K = 3$ for a 99.7% confidence interval, respectively), on a per-user basis. Integrity faults will fall under the "measurement not authenticated" case, and discernment between integrity faults and spoofing is left out of scope of this work. In order to define $\sigma_{auth}$ we need to model the a priori error contributions of all components of $|\tau_{j,E6}^k - \hat{\tau}_{j,E6}^k|$, which can be expressed as a sum of independent error contributions that we here model as Gaussian, zero-mean, and independent. In particular, multipath in E1 and E6 have different envelopes and are modelled independently, although they may be dependent, and also non-zero mean for short durations of time. The selected variance should be large enough to account for this, as well as the environment. With these assumptions, $|\tau_{j,E6}^k - \hat{\tau}_{j,E6}^k|$ can be modeled as a zero-mean Gaussian distribution with variance $\sigma_{auth}^2$, $N(0, \sigma_{auth}^2)$, where

$$\sigma_{auth}^2 = \sigma_{HWB_{rx,E1,E6}}^2 + \sigma_{BGD_{E1,E6}^k}^2 + \sigma_{I,E1}^2 \left(\frac{f_1^2}{f_6^2} - 1\right) + \sigma_{MP,E6}^2 + \sigma_{MP,E1}^2 + \sigma_{n,E6}^2 + \sigma_{n,E1}^2 \quad (25)$$

Annex A develops (25) in more detail, and a proposal for error variance characterization is already presented in [29].

**Further Considerations**

In the proposed approach, no E6C position is calculated. This can be done, but it requires to incorporate time offsets between ECS, when they are not synchronized, in addition to the different time of arrival. Also, the computation of $T_{max}^k$ and $F_{max}^k$ and impact of the E1 pseudorange measurement generation rate are left out of the scope.

Full protection against replay attacks may require to look for power measures (e.g. AGC) or vestigial ECS during the uncertainty time in the receiver. This process is not described in this work, which assumes that if the ECS is found, it is not a replay. Neither is described the logic when the measurement is not authenticated ($\xi_j^k = 0$), which, depending on other factors, may be considered as an indicator of signal spoofing, or just degraded conditions. In any case, when the measurements are authenticated ($\xi_j^k = 1$), which hopefully will be most of the time, and under the assumptions stated, this can be used to ascertain the authenticity of the receiver position, which is the final goal of GNSS authentication.

Finally, one potential drawback of semi-assisted authentication with respect to other methods is that the receiver needs to store the RECS beforehand for the desired period of autonomous operation. Further work may present in more detail receiver storage requirements for sample ACAS configurations, including different autonomy periods and time between authentications.

**CONCLUSIONS AND FURTHER WORK**

A semi-assisted authentication concept proposed for Galileo ACAS (Assisted Commercial Authentication Service) has been described. It is based on the re-encryption of sequences of the E6C, assumed to be encrypted in the near future, their provision a priori to a user receiver over a ground channel at the beginning of the receiver operation, and the a posteriori decryption during operation once the Galileo OSNMA keys are disclosed. The purpose of this approach is to provide signal authentication without any modification to the GNSS signal plan, satellite payload, or signal encryption key management, and without the disclosure of private keys. Yet, the receiver needs to be loosely synchronized with a time source, which is already required for OSNMA.

We describe ACAS in detail, including the re-encrypted sequence (RECS) input files and the E1-E6 broadcast group delays (BGDs). The cryptographic operations to be performed in the receiver are also described. These include the generation of the RECS decryption key from the OSNMA key, the computation of the random location offsets of the RECS, and the decryption of the RECS into the encrypted code sequence, or ECS.

An implementation of the ECS signal correlation is presented. It is based on the storage of an E6C signal snapshot for all satellites, and the adjustment of the snapshot to be correlated for each satellite, once the RECS decryption key is disclosed. In this proposal, the correlation samples are adjusted based on the a-priori synchronization of the receiver to the E1 signal, whose authenticity is verified with the ECS correlation. The receiver accounts for the satellite BGD (available a priori from the ACAS files), the receiver hardware bias between the E1 tracking and the E6 snapshot, to be calibrated, and the frequency dispersive effect of the ionospheric delay. Taking into account these parameters, the receiver can perform the E6 snapshot correlation in a very narrow window in both the time and frequency domains, obtaining an ECS-based E6C pseudorange measurement.

In order to authenticate the E1 position, an authentication verification is proposed whereby the consistency of the E6C and E1 pseudoranges is assessed, defining a threshold based on the a-priori variances of the error contributions. Some aspects not detailed in this work are the authentication status definition in case the signal cannot be authenticated; the combination with other checks from other steps, such as the E6C processing gain, or other sensors; and the verification of vestigial RECS in the receiver loose time interval, to protect against replay attacks. Apart from the abovementioned, further work will also include the verification with real and simulated signals, and the definition of working points for the RECS length and frequency, taking into account real receiver constraints.

**Acknowledgements**

The authors would like to acknowledge the ESA/EUSPA/JRC teams involved in the internal review of the concept, the NACSET and PAULA teams involved in its early prototyping, and G. Caparra and F. Ardizzon for the work and discussions on the topic.

# Annex A – Computation of $\sigma_{auth}$

$\sigma_{auth}$ is the standard deviation of the statistical distribution of $\tau_{E6}^k - \hat{\tau}_{E6}^k$. It is modelled as follows (note if this expression is generalized to any frequencies $f_a$ and $f_b$, the frequency coefficient must be expressed as an absolute value $|\frac{f_a^2}{f_b^2} - 1|$ ):

$$\tau_{E6}^k - \hat{\tau}_{E6}^k = \tau_{E6}^k - \tau_{E1}^k - \delta_{E1,E6}^k =$$

$$\frac{r^k}{c} + \delta t_{rx,E1} + HWB_{rx,E1,E6} - \delta t_{sat,E1}^k + BGD_{sat,E1,E6}^k + (\hat{I}_{E1} + \epsilon_{I,E1})\frac{f_1^2}{f_6^2} + \hat{T} + \epsilon_T + \epsilon_{MP,E6} + \epsilon_{n,E6} -$$

$$(\frac{r^k}{c} + \delta t_{rx,E1} \qquad\qquad - \delta t_{sat,E1}^k \qquad\qquad + \hat{I}_{E1} + \epsilon_{I,E1} \qquad + \hat{T} + \epsilon_T + \epsilon_{MP,E1} + \epsilon_{n,E1}) -$$

$$(\widehat{BGD}_{sat,E1,E6}^k + \hat{I}_{E1}\left(\frac{f_1^2}{f_6^2} - 1\right) + \widehat{HWB}_{rx,E1,E6}) =$$

$$HWB_{rx,E1,E6} - \widehat{HWB}_{rx,E1,E6} + BGD_{sat,E1,E6}^k - \widehat{BGD}_{sat,E1,E6}^k + \epsilon_{I,E1}\left(\frac{f_1^2}{f_6^2} - 1\right) + \epsilon_{MP,E6} - \epsilon_{MP,E1} + \epsilon_{n,E6} - \epsilon_{n,E1} =$$

$$\epsilon_{HWB_{rx,E1,E6}} + \epsilon_{BGD_{E1,E6}^k} + \epsilon_{I,E1}\left(\frac{f_1^2}{f_6^2} - 1\right) + \epsilon_{MP,E6} - \epsilon_{MP,E1} + \epsilon_{n,E6} - \epsilon_{n,E1}$$